\pdfoutput=1 

\documentclass[12pt]{article}
\usepackage{geometry} 
\date{}
\usepackage{subfig}
\usepackage[toc,page]{appendix}
\usepackage{cite}
\usepackage{array}
\usepackage[hidelinks]{hyperref}
\usepackage{graphicx, caption}
\usepackage{enumitem} 
\usepackage{amssymb,amsfonts}
\usepackage{algorithmic}
\usepackage{graphicx}
\usepackage{svg}
\usepackage[strict]{changepage}
\usepackage{xcolor}
\usepackage{textcomp}
\usepackage{siunitx}
\usepackage[T1]{fontenc}
\usepackage[utf8]{inputenc}
\usepackage{float}
\usepackage{array}
\usepackage{booktabs}
\usepackage{adjustbox}

\expandafter\let\csname equation*\endcsname\relax

\expandafter\let\csname endequation*\endcsname\relax
\usepackage{amsmath}

\begin{document}

\author{N. Riva\textsuperscript{1}, F. Sirois\textsuperscript{2}, C. Lacroix\textsuperscript{2}, F. Pellerin \textsuperscript{2}, J. Giguere \textsuperscript{2}, F. Grilli\textsuperscript{3} and B. Dutoit\textsuperscript{1}}

\author{N. Riva\textsuperscript{1}, F. Sirois\textsuperscript{2}, C. Lacroix\textsuperscript{2}, F. Pellerin \textsuperscript{2},\\ J. Giguere \textsuperscript{2}, F. Grilli\textsuperscript{3} and B. Dutoit\textsuperscript{1}\\
$^1${\small École Polytechnique Fédérale de Lausanne, Switzerland} \\
$^2${\small Polytechnique Montréal, Canada} \\
$^3${\small Karlsruhe Institute of Technology, Germany}
}
\title{The eta-beta model model: an alternative to the power-law model for numerical simulations of REBCO tapes}

\maketitle

{
}

\begin{abstract}
When modeling superconducting devices based on REBCO tapes and working near or above the critical current value (i.e. $I>I_{\rm{c}}$), the power-law model is not always accurate. In our previous works, we proposed the overcritical current model, based on a combination of fast pulsed current measurements and finite element analysis. The overcritical current model was provided in the form of look-up tables and was validated experimentally. We showed that the overcritical current model could better reproduce experimental measurements than the power-law model, and that the power-law model predicts a faster quench than the overcritical current model. In this contribution, we propose the eta-beta model, a mathematical expression to model analytically the overcritical current model, based on measurements performed between \SI{77}{\kelvin} and \SI{90}{\kelvin} in self-field conditions. The proposed model is verified by comparing DC fault measurements with the results of numerical simulations using the eta-beta model to represent the electrical resistivity of the superconducting layer of REBCO tapes.
\end{abstract}

\section{Introduction}
\label{sec:introduction}
High-temperature superconductor (HTS) materials in the form of Rare Earth Barium Copper Oxide (REBCO) commercial tapes hold immense potential for future applications such as hadron therapy \cite{Felcini2020}, nuclear fusion \cite{Sorbom2015}, particle accelerators \cite{Nugteren2016}, superconducting fault current limiters (SFCL) \cite{TixadorBook} and superconducting motors \cite{Patel2015}. It is paramount to develop accurate numerical methods for modeling their electro-magneto-thermal behavior. The non-linear $E-J$ relationship of the REBCO material is usually described using the well-established power-law model \cite{Rhyner1993}.

One of the significant drawbacks of the power-law model is that it poorly describes the overcritical current regime ($I>I_{\rm{c}}$) of the superconducting material, and other models remain empirical \cite{Duron2004,Liang2015,Czerwinski2014,9040578,DESOUSA201497,Song_2015}: for instance, they use a piecewise power-law in which the overcritical current regime is modeled  as the flux creep region, but with a lower $n$-value \cite{Paul2001}. Accurate knowledge of the $E-J$ relation can help improve the reliability of the models, and thus, the performance of the simulated devices working in the overcritical current regime or where the electro-magneto-thermal aspects must be studied carefully, for instance a hotspot scenario \cite{Sousa2015,Sousa2016,Bonnard}, a non-insulated coil \cite{Wang2015,Wang2017} or a superconducting dynamo \cite{Mataira2019}.

In our previous works, we developed the \textit{overcritical current model} $\rho_{\rm{{OC}}}(I,T)$, based on fast pulsed current measurements and finite element analysis \cite{Richard2019,Riva2019}. 
We showed that the overcritical current model better reproduces the experimental measurements than the power-law model, and that the power-law model predicts a faster quench than the overcritical current model \cite{Riva2020}. The overcritical current model was provided in the form of a look-up table obtained from data regularization \cite{Nicholson2019,Nicholson2020}. The major drawback of look-up tables is that they are limited to the interval of available data. An extrapolation (or regularization) to a wider data range is not simple to perform.

The model proposed in this work, named eta-beta model, uses a mathematical relationship that overcomes this problem. The mathematical expression of the overcritical current model is based on measurements performed between \SI{77}{\kelvin} and \SI{90}{\kelvin}, in self-field conditions.

The paper is organized as follows. In Section~\ref{sec:experiment}, we outline the experimental methods and the characterization of the samples. In Section~\ref{sec:Dev}, we present the mathematical expression of the overcritical current model, we report the experimental data in the overcritical current regime and compare them with the power-law model. 
Section~\ref{sec:Scale} provides a discussion on the fitting parameters for the overcritical current model and their typical values. 
In Section~\ref{sec:ver}, the overcritical current model is validated by comparing DC fault measurements performed on silver-stabilized samples with the results of a 2-D electro-thermal model implemented in COMSOL Multiphysics.

\section{Experimental methods and characterized}
\label{sec:experiment}
The experimental data in the overcritical current regime were obtained by combining ultra-fast pulsed current measurements and finite element analysis techniques developed in \cite{Sirois2009,Sirois2010,Richard2019,Riva2019}. In the present work, we characterized state-of-the-art REBCO coated conductors from various manufacturers.
We present the results for selected samples fabricated by SuperPower Inc. \cite{SuperPower}, SuperOx \cite{SuperOx}, SuNAM \cite{SuNAM}, and Shanghai Superconductor \cite{Superconductor}; we will refer to these samples as SP01 and SP02 (SuperPower), SO01 and SO02 (SuperOx), SU01 (SuNAM) and SH01 (Shanghai), respectively.

Current pulses up to $I=4\,I_{\rm{c}}$ and as short as \SI{20}{\micro\second} were applied without destroying the samples. The detailed characteristics of the samples are given in Table~\ref{tab:Samples}.
\begin{table}[!htbp]\centering
\begin{tabular}{lcccc}
 \toprule
 \multicolumn{5}{c}{Samples characteristics} \\
 \midrule
 Sample & Width & Silver thickness & REBCO thickness & Hastelloy thickness \\
 & $w_{\rm{tape}}$ & $h_{\rm{Ag,TOT}}$ & $h_{\rm{REBCO}}$ & $h_{\rm{Hast}}$\\
 \hline
SP01/SP02   & \SI{4}{\milli\meter}     &\SI{2}{\micro\meter}/\SI{2.2}{\micro\meter} & \SI{1}{\micro\meter} & \SI{50}{\micro\meter} \\
SO01   & \SI{12}{\milli\meter}     & \SI{4}{\micro\meter} & \SI{6}{\micro\meter} & \SI{97}{\micro\meter} \\
SO02   & \SI{4}{\milli\meter}     & \SI{3}{\micro\meter} & \SI{1}{\micro\meter} & \SI{50}{\micro\meter} \\SU01   & \SI{4}{\milli\meter}     & \SI{4}{\micro\meter} & \SI{1}{\micro\meter} & \SI{50}{\micro\meter} \\
SH01   & \SI{4}{\milli\meter}     & \SI{2.2}{\micro\meter} & \SI{1}{\micro\meter} & \SI{50}{\micro\meter} \\
  \midrule
   \midrule
 \multicolumn{5}{c}{} \\
 \hline
  Sample & Length & Critical current & Critical temperature &  $n$-value \\
  & $l_{\rm{tape}}$ & $I_{\rm{c,\SI{77}{\kelvin}}}$, self-field & $T_{\rm{c}}$ & \\
  \hline
 SP01 & \SI{10.0}{\centi\meter} & \SI{57}{\ampere} & \SI{88}{\kelvin} & $25$\\
 SP02 & \SI{10.0}{\centi\meter} & \SI{110}{\ampere} & \SI{90}{\kelvin} & $30$\\
 SO01 & \SI{9.5}{\centi\meter} & \SI{230}{\ampere} & \SI{92}{\kelvin} & 30\\
 SO02 & \SI{9.5}{\centi\meter} & \SI{177}{\ampere} & \SI{90}{\kelvin} & 30\\
 SU01 & \SI{9.5}{\centi\meter} & \SI{241}{\ampere} & \SI{90}{\kelvin} & 32 \\
 SH01 & \SI{9.5}{\centi\meter} & \SI{177}{\ampere} & \SI{92}{\kelvin} & 28\\
 \bottomrule
\end{tabular}
\caption{The critical current and $n$-value were measured in self-field conditions at \SI{77}{\kelvin} using DC transport current measurements. The critical temperature was determined by performing a resistance vs temperature $R_{\rm{Tape}}(T)$ measurement.}
\label{tab:Samples}
\end{table}

\section{The eta-beta model}
\label{sec:Dev}
\begin{figure}[!htbp]%
    \centering
    \includegraphics[width=36pc]{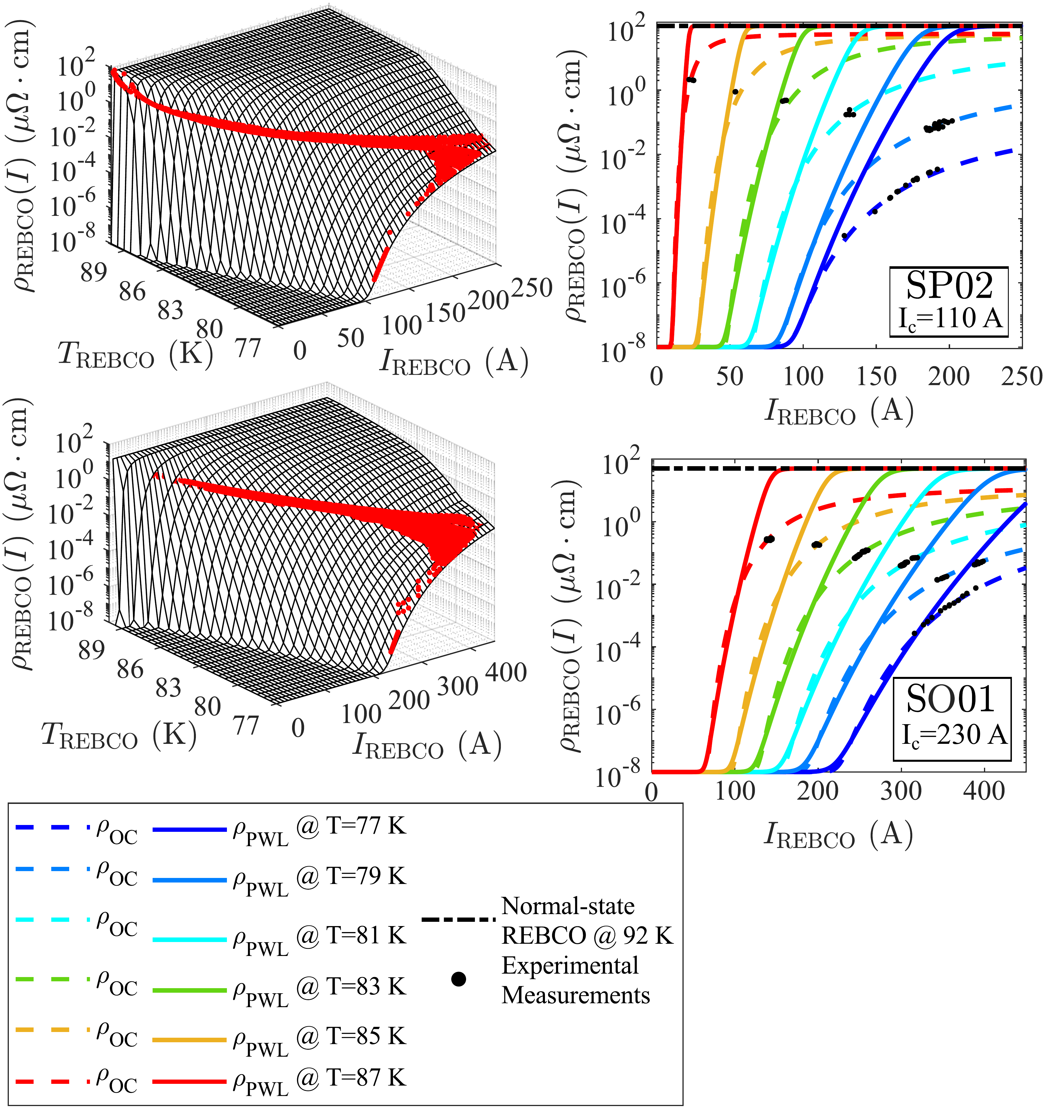}
    \caption{\label{fig:regVSfitSelfField} Results of the modeling for samples SP01 and SO01. We obtained similar results with other samples not shown here. (left) Resistivity surfaces obtained with the $\rho_{\eta\beta}(I,T)$ model. The scattered dots correspond to experimental points obtained using PCM and post-treated using the UC model. (right) Comparison between the experimental measurements (black dots), the $\rho_{\eta\beta}(I,T)$ model (dashed lines) and $\rho_{\rm{PWL}}(I,T)$ model (continuous lines) for selected temperatures.}
\end{figure}

To reproduce the dissipative behavior in the overcritical current regime, we propose a mathematical fitting model. We obtained the model proposed in this work (eta-beta model) by using the Uniform Current (UC) model on pulsed current measurements, presented in figure~\ref{fig:regVSfitSelfField}. The UC model is a post-processing method (thermal simulation plus current sharing model) that allows correcting heating effects from the measurements, hence estimating the temperature, the current sharing and the REBCO resistivity \cite{Richard2019,Riva2019}. The eta-beta model has the following form:
\begin{equation}
    \rho_{\eta\beta}^{\rm{SC}}(I,T)=\rho_{\rm{min}}+\rho_{\rm{c}}\exp{\Bigg[\eta(T)\cdot\Bigg( 1-\bigg[\dfrac{I_{\rm{c}}(T)}{I}\bigg]^{\beta(T)}\Bigg)}\Bigg],
    \label{eq:etabeta}
\end{equation}
where $\rho_{\rm{min}}=10^{-17}\,\si{\ohm\meter}$ (added to avoid numerical problems when the current is~$\sim\SI{0}{\ampere}$), $\rho_{\rm{c}}=10^{-15}\,\si{\ohm\meter}$, $I_{\rm{c}}(T)$ is the temperature-dependent critical current, and $\eta(T)$ and $\beta(T)$ the temperature-dependent fitting parameters. 

Equation~\eqref{eq:etabeta} is based on the mathematical expression of \textit{collective pinning} theory reported in \cite{Larkin1979} and discussed in \cite{Falorio_2014,FALORIO20121462}. The collective pinning theory describes the flux pinning mechanism in single crystal and high quality REBCO films \cite{Falorio_2014,FALORIO20121462}.
At this stage, however, we did not investigate whether the physical quantities described by the collective pinning theory, namely the pinning potential and the glass exponent \cite{Larkin1979}, are correlated to the fitting parameters $\eta(T)$ and $\beta(T)$ of equation~\eqref{eq:etabeta}. The reader should consider equation~\eqref{eq:etabeta}, especially $\eta(T)$ and $\beta(T)$, simply as a mathematical fit that describes the overcritical current regime.

We evaluate the goodness of the mathematical fit with the \textit{normalized root-mean-square deviation} (NRMSD).
We calculated the NRMSD of the fitted resistivity with respect to the experimental data model at various selected temperatures with \cite{JamesWittenBook}:
\begin{equation*}
    \rm{NRMSD}_{\eta\beta}(T)=\dfrac{\sqrt{\dfrac{\sum_{j=i}^{N_{\rm{meas}}} \Big(\rho_{\eta\beta}^{\rm{SC}}(I,T)-\rho_{\rm{UC}}(I,T)\Big)^2}{N_{\rm{meas}}}}}{\rho_{\rm{UC},max}(T)-\rho_{\rm{UC},min}(T)},
\end{equation*}
where $\rho_{\eta\beta}^{\rm{SC}}(I,T)$ and $\rho_{\rm{UC}}(I,T)$ are the eta-beta model and the experimental data calculated at $(I,T)$, respectively, N$_{\rm{meas}}$ is the number of data, and $\rho_{\rm{UC},max}(T)$ and $\rho_{\rm{UC},min}(T)$ are the maximum/minimum values of resistivity obtained from the UC model for the selected temperature $T$. The values of $\rm{NRMSD}_{\eta\beta}(T)$ range between $1$ and $ 12 \%$ for temperatures between \SIrange{77}{90}{\kelvin}.

For purpose of comparison, we also use the power-law model with a temperature-dependent critical current $I_{\mathrm{c}}(T)$. The power-law relationship is written as follows:
\begin{equation}
    \rho_{\rm{PLW}}^{\rm{SC}}(I,T)=\rho_{\rm{min}}+\frac{\Omega\cdot E_{\rm{c}}}{I_{\rm{c}}(T)}\bigg(\frac{|I|}{I_{\rm{c}}(T)}\bigg)^{n-1},
    \label{eq:PWLcurrSCOnly}
\end{equation}
where $\Omega$ is the cross section of the REBCO layer, $E_{\rm{c}}=\SI{1}{\micro\volt\per\centi\meter}$ is the electric field criterion and $n$ is the constant power-law exponent. The critical current value $I_{\rm{c}}(T)$ in both models decreases linearly with the temperature with a relationship of the form:
\begin{equation}
    I_{\rm{c}}(T)=I_{\rm{c,\SI{77}{\kelvin}}}\cdot\Bigg(\frac{T_{\rm{c}}-T}{T_{\rm{c}}-\SI{77}{\kelvin}}\Bigg),
    \label{eq:CurrCritTemp}
\end{equation}
where $I_{\rm{c,\SI{77}{\kelvin}}}$ is the critical current at \SI{77}{\kelvin} in self-field conditions, and $T_{\rm{c}}$ is the critical temperature \cite{Lacroix_2014}.

Finally, for both the eta-beta and the power-law models, the normal state resistivity of REBCO $\rho_{\rm{NS}}(T)$ is added in parallel to that of the superconductor \cite{Duron2004}, i.e.:
\begin{equation}
    \rho_{\rm{k}}=\frac{\rho_{\rm{k}}^{\rm{SC}}(I,T)\cdot \rho_{\rm{NS}}(T)}{\rho_{\rm{k}}^{\rm{SC}}(I,T)+\rho_{\rm{NS}}(T)},
    \label{eq:rhoTotPWLNS}
\end{equation}
where $k$ indicates the two models, namely the power-law model ($\rho_{\rm{PWL}}$) and the overcritical current model ($\rho_{\rm{\rm{\eta\beta}}}$). The temperature dependence of the normal-state resistivity of REBCO is modeled with a simple linear relationship, i.e.: 
\begin{equation*}
    \rho_{\rm{NS}}(T)=\rho_{T_{\rm{c}}}+\alpha\cdot(T-T_{\rm{c}}),
\end{equation*}
where $\rho_{T_{\rm{c}}}$ ranges from \SIrange{30}{100}{\micro\ohm\centi\meter} and $\alpha=\SI{0.47}{\micro\ohm\centi\meter\per\kelvin}$ according to the literature \cite{Friedmann,Bonnard}.

In figure~\ref{fig:regVSfitSelfField} (left) we present the raw overcritical current data (red-scattered dots) extracted with the UC model (see \cite{Richard2019,Riva2019}) and the resistivity surface obtained with the $\rho_{\eta\beta}(I,T)$ model.
All the measurements start at \SI{77}{\kelvin}, the temperature of the liquid nitrogen bath, where the samples were immersed during the characterization.
The maximum value of the normal state REBCO resistivity ($\rho_{\rm{NS}}$) calculated with the UC model ranges from \SIrange{50}{100}{\micro\ohm\centi\meter}, in agreement with the literature for $\rho_{\rm{NS}}$ above the critical temperature $T_{\rm{c}}$ \cite{Bonnard,Friedmann}. The maximum resistivity corresponds to temperatures between \SI{88}{\kelvin} and \SI{88}{\kelvin}, in agreement with the typical values found for REBCO \cite{Wu1987}.
In figure~\ref{fig:regVSfitSelfField} (right) we compare, for selected temperatures, the raw overcritical current data (scattered black-dots) with the overcritical current model $\rho_{\eta\beta}(I,T)$ (dashed lines), and with the power-law model $\rho_{\rm{PWL}}(I,T)$ (continuous lines). 
The difference between the experimental overcritical current data (thus the $\rho_{\eta\beta}(I,T)$ model) and the $\rho_{\rm{PWL}}$ is remarkable. In comparison with the power-law, the overcritical current data and the eta-beta model, present a much reduced slope with respect to the power-law.

Such results corroborate the work done by Falorio et al. \cite{FALORIO20121462,Falorio_2014}, where a similar deviation from the power-law model was observed on four etched samples and across a very large range of temperatures (\SIrange{15}{86}{\kelvin}).

\section{Analysis of the fitting parameters}
\label{sec:Scale}
This section presents a detailed discussion about the fitting parameters $\eta(T)$ and $\beta(T)$ used in the $\rho_{\eta\beta}(I,T)$ model.
\subsection{Fitting parameters in self-field conditions}
\begin{figure}[!htbp]%
    \centering
    \includegraphics[width=36pc]{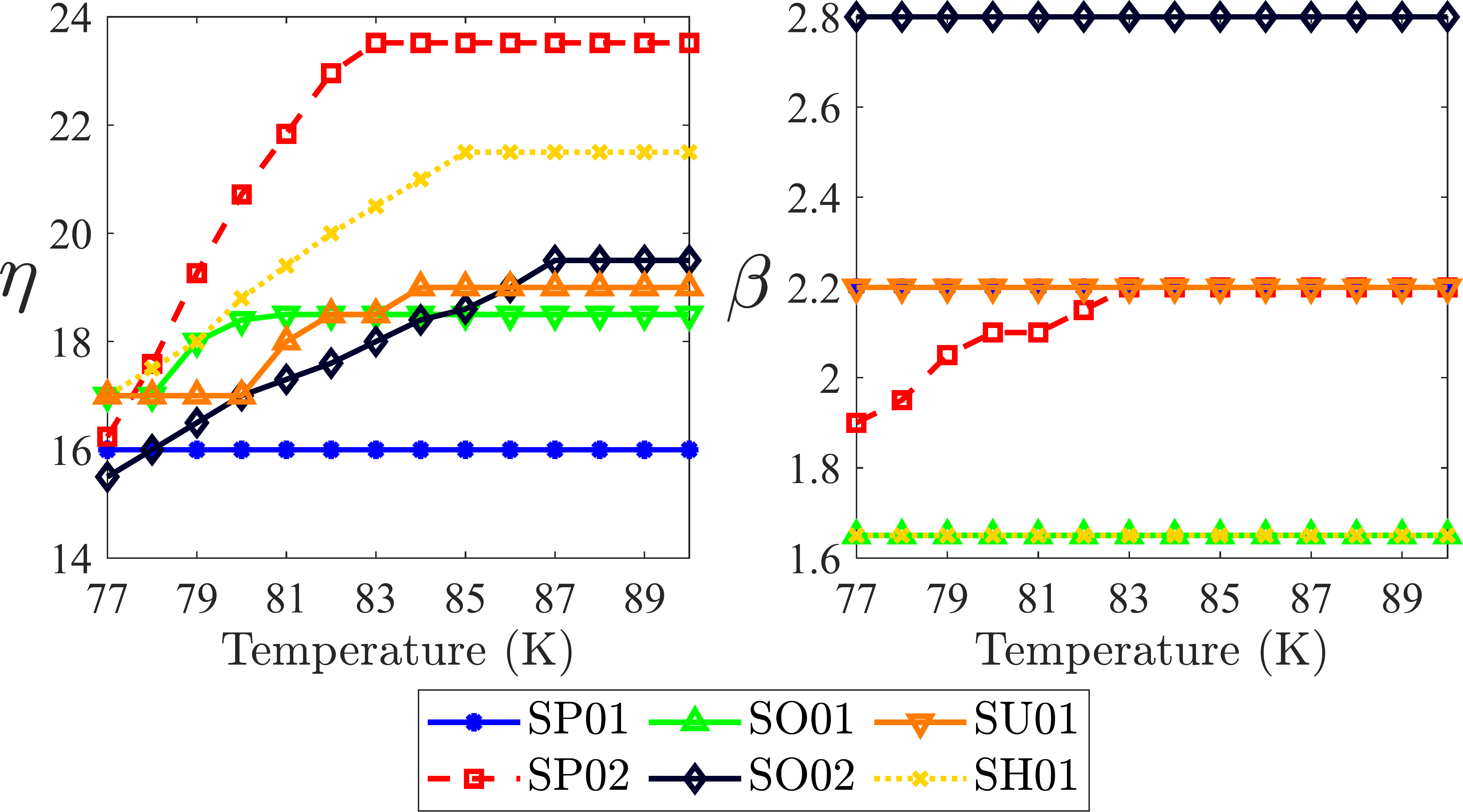}
    \caption{\label{fig:etabetaParamSelfField} Temperature dependence of the fitting parameters (left) $\eta$ and (right) $\beta$ of the $\rho_{\eta\beta}(I,T)$ model for various commercial samples.}
\end{figure}

In figure~\ref{fig:etabetaParamSelfField} we present the temperature dependence of the fitting parameters $\eta(T)$ and $\beta(T)$ for all samples in self-field condition. The data show similarities for all samples.
Sample SP01 presents a constant $\eta(T)$ at all temperatures. For all other samples, $\eta(T)$ increases in a monotonic way and eventually reaches a plateau. In the case of samples SO01 and SUO1, $\eta(T)$ remains constant until a specific value of temperature (\SI{78}{\kelvin} and \SI{80}{\kelvin}, respectively) before it starts to increase. The parameter $\beta(T)$ is constant for most of the samples between \SI{77}{\kelvin} and \SI{90}{\kelvin}, except for sample SP02.

We can fit a piecewise polynomial function of $2^{\rm{nd}}$ and $1^{\rm{st}}$-degree on $\eta(T)$ and $\beta(T)$, respectively. For $\eta(T)$ the relationship can be written as follows:
\begin{equation}
    \eta(T)=\begin{cases} 
    a_{0} & 77\leq T< T_{\rm{1}}\\
    a_{2}(T-T_{\rm{1}})^2+a_{1}(T-T_{\rm{1}})+a_0 & T_{\rm{1}}\leq T< T_{\rm{2}} \\
    \eta_{\rm{max}} & T\geq T_{\rm{2}},
   \end{cases}
   \label{eq:etaFunction}
\end{equation}
where $T$ is the temperature of the REBCO material, $T_{\rm{1}}$ and $T_{\rm{2}}$ are the threshold temperatures below and above which $\eta$ remains constant, and $a_i$ (with $i=0,1,2$) are the polynomial coefficients. The parameter $\eta_{\rm{max}}$ is determined by imposing a continuity condition such that $\eta(T_2)=\eta_{\rm{max}}$.

The piecewise polynomial relationship for $\beta(T)$ is the following:
\begin{equation}
    \beta(T)=\begin{cases} 
      b_{1}(T-T_{\rm{1}})+b_0 & 77\leq T< T_{\rm{2}} \\
      \beta_{\rm{max}} & T\geq T_{\rm{2}},
   \end{cases}
   \label{eq:betaFunction}
\end{equation}
where $T$ is the temperature of the REBCO material, $T_{\rm{2}}$ a threshold temperature above which $\beta$ remains constant, and $b_i$ (with $i=0,1$) are the polynomial coefficients.

The parameter $\beta_{\rm{max}}$ is determined by imposing a continuity condition such that $\beta_{\rm{max}}=\beta(T_2)=b_{1}(T_{\rm{2}}-T_{\rm{1}})+b_0$.

Table~\ref{tab:FittingParamTable} contains a summary of the fitting parameters for all samples. For sample SP02, we tried both a quadratic and a linear fit (SP02 and SP02$_{\rm{lin}}$, respectively). \\
\begin{table}[!htbp]\centering
  \setlength\tabcolsep{5pt}
\begin{tabular}{l|cccc|ccc|cc|cc}
 \toprule
 \multicolumn{12}{c}{Fitting parameters} \\
 \hline
  Sample & $a_2$ & $a_1$ & $a_0$ & $\eta_{\rm{max}}$ & $b_1$ & $b_0$ & $\beta_{\rm{max}}$ & $T_{\rm{1}}$ & $T_{\rm{2}}$ & $T_{\rm{c}}$  & $\rho_{\rm{NS}}$\\& [\si{\per\kelvin\squared}] & [\si{\per\kelvin}] & [1] & [1] & [\si{\per\kelvin}] & [1] & [1] & [\si{\kelvin}] & [\si{\kelvin}] & [\si{\kelvin}]& [\si{\micro\ohm\centi\meter}]\\
  \hline
 SP01 & - & - & 16.00 & 16.00 & - & - & 2.20 & - & - & 88 & 100\\
 SP02 &  -0.0170 & 1.3340 & 16.23 & 23.62 & 0.048 & 1.92 & 2.20 & 77 & 83 & 90 & 100\\
 $\rm{SP02_{\rm{lin}}}$ & - & 1.2331 & 16.23 & 23.62 & 0.048 & 1.92 & 2.20 & 77 & 83 & 90& 100\\
 S001 & -0.2360 & 1.2100 & 17.00 & 18.50 & - & - & 1.65 & 78 & 81& 92 & 79\\
  S002 & - & 0.3800 & 15.68 & 19.48 & - & - & 2.80 & 77 & 87& 90& 53\\
SU01 & -0.1364 & 1.0360 & 17.03 & 19.00 & - & - & 2.20 & 80 & 84& 90& 54\\
 SH01 & - & 0.6150 & 17.00 & 21.92 & - & - & 1.65 & 77 & 85& 92& 89\\
 \bottomrule
\end{tabular}
\caption{Fitting parameters of the $\rho_{\eta\beta}(I,T)$ model, critical temperature $T_{\rm{c}}$ (measured) and normal state resistivity calculated with the UC model at $T_{\rm{c}}$ for each sample.}
\label{tab:FittingParamTable}
\end{table}

\section{Validation of the $\rho_{\eta\beta}(I,T)$ model: DC limitation tests vs numerical simulations}
\label{sec:ver}
\subsection{Critical current measurements and DC limitation tests\label{sect:DCtest}}
\begin{figure}[!htbp]
    \centering
   \includegraphics[width=3in]{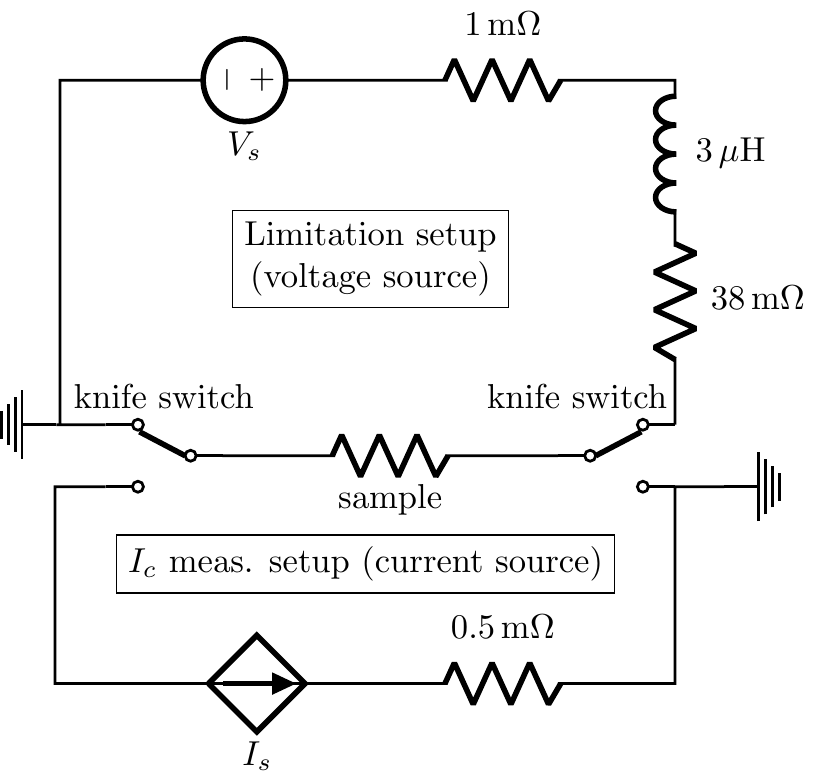}
    \caption{Simplified illustration of the experimental setup using two custom-made sources for performing limitation tests and critical current measurements. }
    \label{fig:SchemaElect}
\end{figure}

DC current limitation are used in this paper to validate the developed model. The DC current limitation test setup is schematically represented in figure~\ref{fig:SchemaElect}. It uses two custom-made sources: a voltage source $V_{\rm{s}}$ and a current source $I_{\rm{s}}$. In both cases, the electric energy is provided by a large supercapacitor (\SI{48}{\volt}/\SI{165}{\farad}). The voltage source was used for the limitation tests, while the current source was used for critical current measurements. The sample can be connected to either one of the two sources using a knife switch (dual pole, dual throw). Two shunts with values of \SI{1}{\milli\ohm} and \SI{0.5}{\milli\ohm} were used to monitor the current generated by the voltage and the current sources, respectively. 

\begin{figure}[!htbp]
    \centering
    \includegraphics[width=3in]{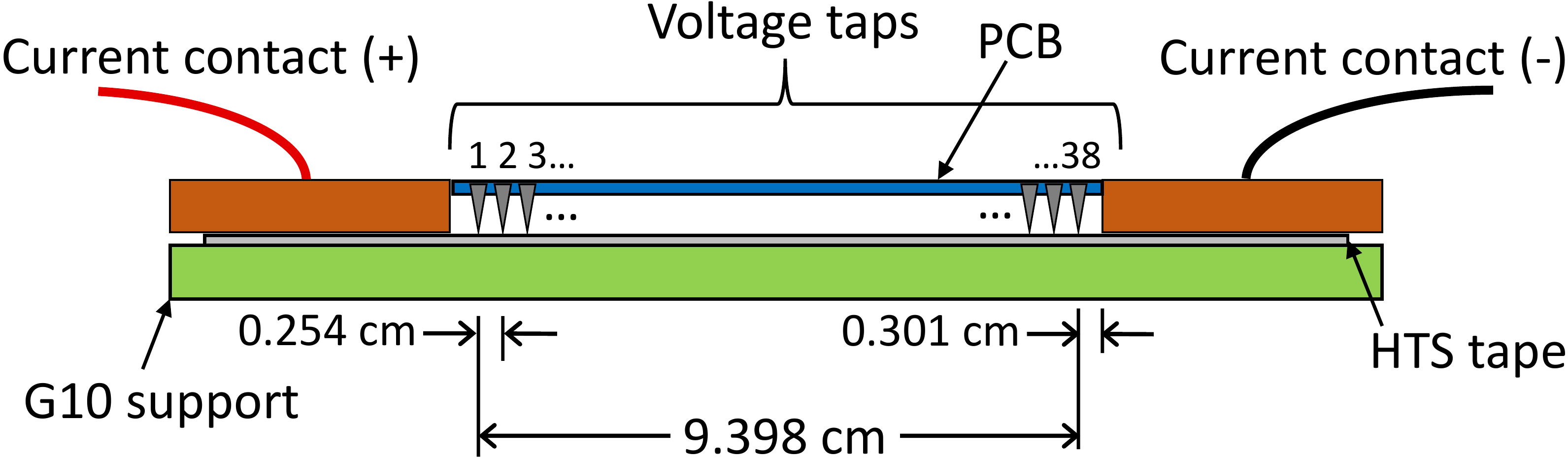}
    \caption{Illustration of the sample holder used for electrical characterization. Thirty eight pogo pin probes (voltage taps) were used to monitor the voltage drop in different sections of the HTS tape. The numbers on top of each voltage tap are used to identify them. }
    \label{fig:SampleHolder}
\end{figure}

The sample holder used for the electrical measurements allowed testing 14-cm-long samples and had 38 evenly spaced spring-loaded contacts (\SI{2.54}{\milli\meter} spaced pogo pins), identified as voltage taps in figure~\ref{fig:SampleHolder}. The voltage taps are used to measure the voltage drop along the sample length. The pogo pins were soldered on a custom-made PCB, which was screwed in a G10 support to adjust the pressure of the pogo pins on the sample. The current leads were made of thick copper blocks that were also screwed to the G10 support. To minimize the contact resistance at the current contacts, a thin indium sheet was inserted between the HTS tape and the copper blocks at each end of the sample. Finally, additional voltage taps were attached to both current contacts in order to monitor the contact resistance during the measurements. 

During the experiments, the sample and sample holder were immersed together in a liquid nitrogen bath (\SI{77}{\kelvin}). Prior to each limitation test, the current source circuit was used to measure the critical current of the sample, using 10-ms-long square current pulses. The total voltage drop between pin 1 and 38 was measured. A National Instruments card (NI-8255) was used for the data acquisition. For each pulse, the current amplitude was increased until the critical electric field criterion was reached (\SI{1}{\micro\volt\per\centi\meter}). The measured critical current for sample SP02b, cut from the same spool as SP02, is $I_{\rm{c}}=\SI{109}{\ampere}$ and the $n$-value is $n=24$. After the critical current measurement, the sample was disconnected from the current source and connected to the voltage source to perform a DC limitation test. The voltage $V_{\rm{s}}$ of the supercapacitor and the length of the pulse were set to the desired values. Differential measurements of the voltage drop between two adjacents voltage taps and the voltage drop between the two extreme taps were monitored during each measurement. Then, after the limitation test, the sample was disconnected again from the voltage source and reconnected to the current source in order to re-measure the $I_{\rm{c}}$ and verify if the sample suffered degradation or not during the limitation test. 
\subsection{Numerical model}
\begin{figure}[!htbp]%
    \centering \includegraphics[width=30pc]{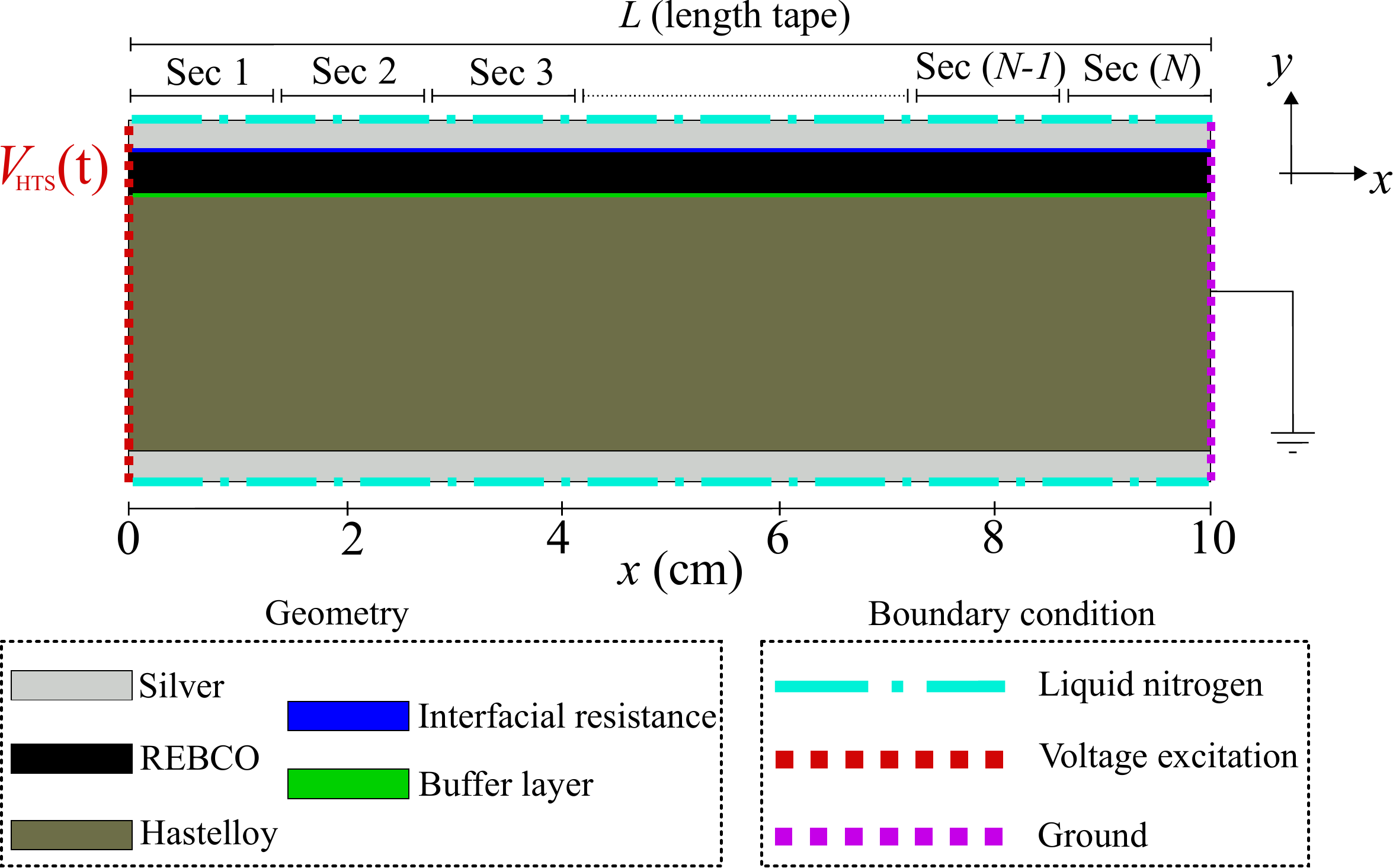}
    \caption{\label{fig:PointProbeSec} Tape geometries used in finite element calculations (not to scale) and boundary conditions.}%
\end{figure}

In this work, we developed a 2-D longitudinal cross-section electro-thermal model in COMSOL Multiphysics \cite{COMSOL}. This means that we considered the material properties to be uniform across the width of the tape. This model allowed us to simulate inhomogeneities and hot-spots along the length of the tape. Figure~\ref{fig:PointProbeSec} is a schematic representation of the 2-D longitudinal cross-section of the tape.
The tape consists of a REBCO layer (black), a MgO buffer layer (green) deposited on a substrate (brown), and a surrounding layer of silver stabilizer (gray). The interfacial resistance layer (blue) between the REBCO and the silver layer is also modeled. The dimensions of each layer of the tape are given in Table~\ref{tab:Dimensions}, which correspond to the dimensions of the SuperPower sample used for the validation of the eta-beta model.
\begin{table}[b]\centering
\begin{tabular}{llr}
 \toprule
 Parameter  & Symbol & Numerical value\\
 \midrule
 Tape length & $L$ & \SI{91.64}{\milli\meter}\\
 Tape width & $w_{\rm{tape}}$ & \SI{4}{\milli\meter}\\
 Silver thickness & $h_{\rm{Ag}}$ & \SI{2.2}{\micro\meter}\\
 Buffer layers thickness & $h_{\rm{MgO}}$ & \SI{150}{\nano\meter}\\
 Interfacial layer thickness (REBCO - Ag) & $h_{\rm{in}}$ & \SI{100}{\nano\meter}\\
 REBCO thickness & $h_{\rm{REBCO}}$ & \SI{1}{\micro\meter}\\ 
 Hastelloy thickness & $h_{\rm{Hast}}$ & \SI{47}{\micro\meter}\\
 \bottomrule
\end{tabular}
\caption{Geometric parameters of REBCO tapes used in the simulations.}
\label{tab:Dimensions}
\end{table}
The state variables used are the electrical potential $V$ and the temperature $T$. The following equation is used to solve for the electrical potential:
\begin{equation}
    \nabla \cdot(-\sigma(I,T) \nabla V(t))=0,
    \label{eq:Gauss}
\end{equation}
where $V(t)$ is the electric scalar potential, and $\sigma(I,T)=1/\rho(I,T)$ is the temperature-dependent electrical conductivity.
We used the measured voltage $V_{\mathrm{HTS}}(t)$ as a voltage excitation applied on the left boundary of the tape (dashed red line), while on the right boundary of the tape (dashed purple line) we imposed a ground ($V=0$). Finally, on the remaining boundaries of the conductor, we have:
\begin{equation}
    \textbf{n}\cdot \nabla V=0.
\end{equation}
The heat equation is solved on the 2-D domain represented in figure~\ref{fig:PointProbeSec} as follows:
\begin{equation}
    \rho_{\mathrm{mass}}(T)C_{\mathrm{p}}(T)\frac{\partial T}{\partial t}+\nabla\cdot\big(-k(T)\nabla T\big)= \sigma(I,T)(-\nabla^2 V ) -h_{\rm{LN_2},(T-T_{\rm{0}})}\cdot (T-T_{\rm{0}})\bigg|_{\partial\Omega},
    \label{eq:HeatEquation}
\end{equation}
where on the left-hand side of the equation, we have the mass density $\rho_{\mathrm{mass}}(T)$, the specific heat capacity $C_{\mathrm{p}}(T)$ and the thermal conductivity $k(T)$. On the right-hand side of the equation, there is the heat source term and a cooling term. The term $\sigma(I,T)(-\nabla^2 V )$ represents the heat source Joule losses, where the voltage gradient $\nabla V$ is calculated from equation~\eqref{eq:Gauss}.
The heat exchange with liquid nitrogen is taken into account by applying a boundary condition applied only on the top and the bottom silver layers of the tape (figure~\ref{fig:PointProbeSec} dashed-dotted sky-blue), indicated by $\partial\Omega$ in equation~\eqref{eq:HeatEquation}. The heat transfer coefficient $h_{\rm{LN_2},(T-T_{\rm{0}})}$ is a function of the temperature \cite{Francois2010}. Finally, the temperature dependencies of all the thermal and electrical material properties are taken into account and can be obtained from the literature \cite{Zou2017,Roy2008,Lacroix_2014}.
For the REBCO layer, we considered both the overcritical current model $\rho_{\eta\beta}(I,T)$ and the power-law model. For sample SP02 we implemented the overcritical current model $\rho_{\eta\beta}(I,T)$ according to the parameters $\eta(T)$ and $\beta(T)$ in Table~\ref{tab:FittingParamTable}. We implemented whe power-law model as in equation~\eqref{eq:PWLcurrSCOnly}-\eqref{eq:rhoTotPWLNS} with a constant $n$-value of $n=30$.

Finally, we toke into account of the critical current inhomogeneities along the length of the tape by subdividing the tape in $N$ equal sections, of length $L_{\rm{sec}}=L/N$ (see figure~\ref{fig:PointProbeSec}), and by explicitly specifying the value of the local critical current in each section.
The local critical current values are specified using a spatial function $\Phi(x)$ multiplied by the nominal critical current $I_{\rm{c,nom}}(T)$:
\begin{equation}
    I_{\rm{c}}(T,x)=\Phi(x)\cdot I_{\rm{c,nom}}(T).
    \label{eq:CurrentDistrib}
\end{equation}
Note that if $\Phi(x)=1$ the tape is perfectly homogeneous; if $\Phi(x)=0$ the tape has no superconducting properties.

In our model, $\Phi(x)$ is a piecewise function composed of $N$ constant intervals and representing the $N$ intervals along the tape:
\begin{equation}
        \Phi(x)=
        \begin{cases}
            \Phi_1, & \mathrm{Section\,\,1}\\
            \Phi_2, & \mathrm{Section\,\,2}\\
            ... & ...\\
            \Phi_i, & \mathrm{Section\,\,}i\\
            ... & ...\\
            \Phi_N, & \mathrm{Section\,\,}N
        \end{cases}
        \label{eq:IcInhomogen}
    \end{equation}
The parameters $\Phi_i$ (with $i=1,2,...,N$) are the coefficients that specify the local deviation of the local critical current $I_{\rm{c}}(T,i)$ from the nominal value of $I_{\rm{c,nom}}(T)$. For example, specifying $\Phi_i=1.1$ and $\Phi_j=1.2$ means that the $i$-th and $j$-th regions are such that $I_{\rm{c}}(T,i)=1.1\, I_{\rm{c,nom}}(T)$ and $I_{\rm{c}}(T,j)=1.2\, I_{\rm{c,nom}}(T)$, respectively.

The parameters $\Phi_i$ are \textit{normally distributed} (Gaussian distribution) around a \textit{mean} $\Phi_{\mathrm{av}}$ and with a \textit{standard deviation} $\sigma_{\Phi,\%}$\footnote{Note that if $\sigma_{\Phi,\%}$ goes to 0, the critical current distribution approaches perfect homogeneity.}. 
It is useful to recall that the nominal critical current of REBCO tape $I_{\rm{c,nom}}$ is defined by the critical current measured electrically over the whole tape. For this reason, for a given degree of inhomogeneity, we fixed the minimum value of the critical current distribution to be always the measured critical current $I_{\rm{c,nom}}$. This means that the minimum (or maximum)
of the critical current distribution are within $\pm3\,\sigma_{\Phi,\%}$ of the average critical current $I_{\rm{c},av}(T)$ \footnote{According to the statistic of the normal distribution, the values less than three standard deviation ($\pm 3\,\sigma_{\Phi,\%}$) away from the average ($I_{\rm{c},av}(T)$) account for $99.73\,\%$ of the set\cite{Crilly1997}.}.

The average $I_{\rm{c},av}(T)$ of the critical current distribution $I_{\rm{c}}(T,x)$ instead, depends on the degree of inhomogeneity. For instance, if $I_{\rm{c,nom}}=\SI{109}{\ampere}$, an inhomogeneity degree of $20\,\%$ implies $I_{\rm{c},av}(T)=I_{\rm{c,nom}}/(1-0.2)\simeq\SI{136}{\ampere}$.

By referring to the normally distributed parameters $\Phi_i$ in equation~\eqref{eq:IcInhomogen}, we have $\Phi_{\mathrm{av,20\,\%}}\simeq1.25$, $\sigma_{\Phi,20\,\%}=0.083$, and the minimum of the parameters $\Phi_{i}$ is $\Phi_{\mathrm{min}}=\big(\Phi_{\mathrm{av,20\,\%}}- 3\,\sigma_{\Phi,20\,\%}\big)\simeq 1$.

This means that the average of the critical current distribution $I_{\rm{c},av}(T)$ is:
\begin{align*}
   I_{\rm{c},av,20\,\%}(T)&=\Phi_{\mathrm{av}}\cdot I_{\rm{c,nom}}(T)\simeq\SI{136}{\ampere},
\end{align*}
and the minimum of the critical current distribution corresponds to the measured critical current $I_{\rm{c,nom}}(T)$:
\begin{align*}
   I_{\rm{c,nom}}(T)&=\Phi_{\mathrm{min}}\cdot I_{\rm{c,nom}}(T)=\simeq\SI{109}{\ampere}.
\end{align*}
Finally, we selected $N=30$ sections, which means around \SI{3}{\milli\meter} per section for our setup measurements. In reality, the variation of $I_{\rm{c}}$ occurs at a much smaller scale. We choose of using $N=30$ sections as a compromise, in order to obtain some insights of the influence of $I_{\rm{c}}$ non-uniformity while keeping the computation time reasonable.

\subsection{Comparison with numerical results}
\begin{figure}[!htbp]%
    \centering \includegraphics[width=38pc]{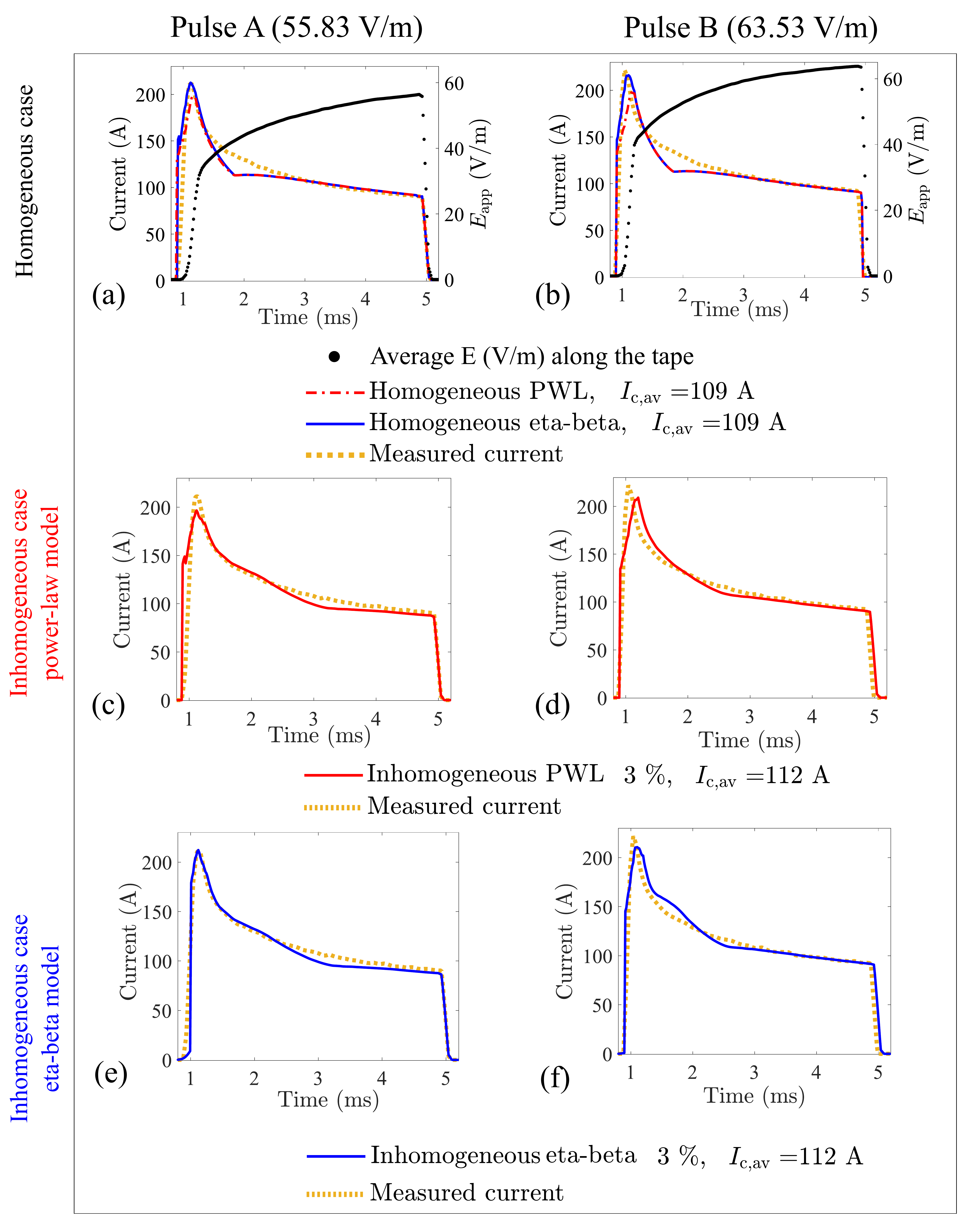}
    \caption{\label{fig:Merge} Measurements and simulations of the SP02b sample undergoing DC current limitation. The experimental current peak at the beginning for Pulse A is $I_{\rm{peak}}=\SI{212.5}{\ampere}$ and for Pulse B is $I_{\rm{peak}}=\SI{219.4}{\ampere}$.}%
\end{figure}

In Figure~\ref{fig:Merge}, we compare the results of the experimental measurements and the simulations for sample SP02b for homogeneous and inhomogeneous critical current distributions.
In figure~\ref{fig:Merge} left (Pulse A), the maximum voltage (electric field) applied on the sample was \SI{5.11}{\volt} (\SI{55.83}{\volt\per\meter}), which generated a peak current $I_{\rm{peak}}=\SI{212.5}{\ampere}$.
In figure~\ref{fig:Merge} right (Pulse B), the maximum voltage (electric field) applied on the sample was \SI{5.82}{\volt} (\SI{63.53}{\volt\per\meter}), which generated a peak current $I_{\rm{peak}}=\SI{219.4}{\ampere}$.
If we consider the simulation results obtained for a homogeneous critical current distribution (figure~\ref{fig:Merge}(a-b)), the measured current profile (yellow) is overall well reproduced, especially the current peak at the beginning when using the $\rho_{\eta\beta}$ model (red-dashed). The $\rho_{\rm{PWL}}$ model (blue) underestimates the initial current peak at the beginning.
Both resistivity models, however, fail around~\SI{2}{\milli\second}. A discrepancy between the DC limitation tests and the simulations was observed already in a previous work of ours \cite{Riva2020}. The abrupt change of slope around \SI{1.8}{\milli\second} in the homogeneous case indicates that the entire tape is quenched and all the current flows in the stabilizer and the Hastelloy.

A possible explanation since the current regime explored in these measurements is near the critical current of the tape (\SI{109}{\ampere}) and the critical current distribution is not uniform, the tape does not quench uniformly and different regions quench at different time.

To verify this hypothesis, we implemented an inhomogeneous critical current distribution, as described in the previous section. For our simulations, we selected an inhomogeneity degree of $3\,\%$, which means $I_{\rm{c,nom}}=\SI{109}{\ampere}$, $I_{\rm{c},av}(T)=I_{\rm{c,nom}}/(1-0.03)\simeq\SI{112}{\ampere}$, $\Phi_{\mathrm{av,3\,\%}}\simeq1.03$ and $\sigma_{\Phi,3\,\%}=0.01$.

For both DC limitation tests, an inhomogeneous critical current distribution improves the matching between the simulated and experimental current profile. These results confirm our previous hypothesis about the impact of the inhomogeneity on the simulated current; an inhomogenoeus critical current distribution improves the matching between simulations and experiments at \SI{2}{\milli\second}. Specifically, since different regions quench at different time, the overall resistance of the tape is lower and the current is higher, which explains the smoother profile of the simulated current.

Further remarks on the simulation results are necessary. Firstly, there are still some discrepancies around \SI{2}{\milli\second}, which are probably due to the difficulty of knowing the exact degree of inhomogeneity and the exact position of the defects in the tape. A precise knowledge of the critical current distribution may further improve the matching between experiments and simulations. However, we verified that simply changing the degree of non-uniformity of $I_{\rm{c}}$ (up to $20\,\%$) does not improve the agreement with experimental results. Secondly, the current rise at the beginning of the simulated current pulse is too sharp with respect to the measured current. This difference may be due to an incorrect compensation of the voltage induced in the voltage tap during the measurements.

The simulation results, however, confirm that the eta-beta model reproduces better the temporal profile of the measured current accurately for the two electric fields level considered, especially the current peak around \SI{1}{\milli\second}.

\section{Conclusions}
\label{sec:conclusion}
In this paper we introduced the eta-beta $\rho_{\eta\beta}(I,T)$ model as an alternative to the power-law model. The advantage of the $\rho_{\eta\beta}(I,T)$ model is its extended range of validity in the overcritical current regime. Through numerical simulations, we validated the $\rho_{\eta\beta}(I,T)$ model with DC fault current measurements. 
In \cite{Riva2020}, it was demonstrated that the power-law model predicts a faster quench than the overcritical current model, but the latter was provided in the form of look-up table. The  $\rho_{\eta\beta}(I,T)$ model is analytic and compact and can be conveniently implemented in numerical simulations for other cases in which current sharing plays an important role (e.g., hot spots in superconducting devices or non-insulated coils). In the future, an extensive measurement campaign will be necessary for exploring several experimental conditions (samples with different pinning force, measurement in magnetic field, wider temperature range, etc.). 
Although we characterized many samples from different manufacturers, we have not provided a physical interpretation of the fitting parameters. This paper opens the study of the overcritical current regime by numerical simulations, a new exciting aspect of REBCO commercial tapes.

\bibliographystyle{unsrt}
\bibliography{Main}
\end{document}